\documentclass[floatfix,showkeys,citeautoscript,showpacs,amsmath,amssymb,preprint]{revtex4}
\usepackage{graphicx}
\usepackage{dcolumn}
\usepackage{amsmath}
\usepackage[dvipsnames]{xcolor}
\usepackage{multirow}
\usepackage{longtable}
\usepackage{makecell,array}
\usepackage{longtable}
\begin{document}

\title{Static dipole polarizabilities of polyacenes using self-interaction-corrected density functional approximations}
\author{Sharmin Akter$^1$}
\author{Yoh Yamamoto$^2$} \author{Rajendra R. Zope$^2$} \author{Tunna Baruah$^2$}
\email{tbaruah@utep.edu}
\affiliation{$^1$Computational Science Program, University of Texas at El Paso, El Paso, TX, 79968}
\affiliation{$^2$Department of Physics, University of Texas at El Paso, El Paso, TX, 79968}

\date{\today}

\begin{abstract}
   Density functional approximations are known to significantly overestimate the polarizabilities of long chain-like molecules.  We study the static electric dipole polarizabilities and the vertical ionization potentials of polyacenes from benzene to pentacene using the Fermi-L\"owdin orbital based self-interaction corrected (FLOSIC) density functional method. The orbital-by-orbital self-interaction correction corrects for the overestimation tendency of density functional approximations. The polarizabilities calculated with FLOSIC-DFA are however overly corrected. We also tested the recently developed locally-scaled self-interaction correction (LSIC) method on the polyacenes. The local-scaling method applies full SIC in the one-electron regions and restores the proper behavior of the SIC exchange-correlation functionals in the uniform density limit. The results show that  LSIC removes the overcorrection tendency of the FLOSIC-DFA and produces results that are in excellent agreement with reference CCSD values. The vertical ionization potentials with LSIC also show good agreement with available experimental values.

\end{abstract}
\maketitle

\section{Introduction}
   
   Static dipole polarizability of an atom  or molecule 
  describes how easily the electron cloud can be deformed under the influence of an external static electric field. When a system is placed in a static electric field, the electronic charge of the system  redistributes. This response can be characterized by induced multipole moments. Of these, the induced dipole moment is related to the static dipole polarizabilities. Static electric dipole polarizabilities measure the change at first order in the molecular dipole moment  with the interaction of external homogeneous  electric field\cite{bonin1997electric}.  

  Polyacenes are two-dimensional conjugated aromatic molecules that can be considered as serially fused benzene rings. The polyacenes form organic semiconductors that can be used in 
  optoelectronic devices such as organic field-effect transistors, solar cells, and light-emitting diodes \cite{tang1987organic, vanslyke1988electroluminescent,burroughes1990light}. 
  The dielectric properties of polyacenes and their derivatives are widely studied for 
  their possible usage in such devices.
  Experimentally, a variety of techniques have been used to measure the polarizabilities  of polyacenes such as laser Stark-effect spectroscopy \cite{heitz1992measurement,bendkowsky2007high}, Kerr-effect experiments \cite{JR9550001641,kuball1969untersuchungen}, Cotton-Moutton-effect \cite{cheng1972molecular}, electric field NMR \cite{ruessink1986anisotropy}, crystal refraction \cite{vuks1966determination}.
  From a theoretical point of view, there are also a numerous number of methods that have been used so far for calculating static dipole polarizabilities of polyacenes such as coupled-cluster single and double linear response (CCSD-LR) theory \cite{hammond2007dynamic}, Hartree-Fock (HF) theory \cite{firouzi2008polyacenes,millefiori1998ab}, density functional theory (DFT) \cite{soos2001charge,reis2000calculation,smith2004static,hinchliffe2005density} with different functionals and basis sets.
  The failure of density functional approximations (DFAs) in the prediction of dipole and higher-order polarizabilities for molecular chains and polymers has been discussed extensively by several groups \cite{champagne_jcp1998,Kirtman,Burke,kummel, PhysRevA.52.178,doi:10.1063/1.4864038,weitao2006,Adrienne_PRA_2008}. The response of approximate 
exchange-correlation (XC) potentials in an applied electric field differs from that of the exact XC potential. The exact XC potential develops a counteracting field whereas with the local density approximation (LDA) or the the generalized gradient approximation (GGA), the induced XC potential acts in the same direction as the applied field \cite{Kirtman,Ortiz1998}. This effect leads to large polarization of the material, particularly in extended systems.

  An underlying problem of the approximate density functionals is the so-called self-interaction error (SIE)
 which was discussed extensively in a seminal paper by Perdew and Zunger \cite{PZSIC81}. 
 Unlike the Hartree-Fock method, the approximate self-exchange 
 energy does not 
 completely cancel out the self-Coulomb energy leading to the unphysical behavior where an electron interacts with itself. The presence of SIE results in the potential decaying as $-exp(-r)$ instead of $-1/r$. The SIE 
 makes the electrons overly delocalized with DFAs which can be seen
 in high values of dipole polarizabilities calculated with DFAs.  Earlier applications of self-interaction correction (SIC) have been mainly on hydrogen molecular chain with various
 SIC methods including optimized effective potential (OEP)\cite{kummel},  with OEP within the Krieger-Li-Iafrate approximation \cite{Burke}, Perdew-Zunger SIC (PZSIC) \cite{Vydrov2008}, or with one-electron SIE free functionals \cite{weitao2006}. These applications have shown significant improvement of the linear polarizability with SIC.

  In this work,  
  we examine the effect of one-electron SIE 
  on static dipole polarizabilities and ionization potentials (IPs) of five polyacenes 
  from benzene to pentacene. These molecules offer the possibility to systematically 
  study the effect of SIE as the system size grows.
  Both experimental and highly accurate theoretical values are also available for these molecules. 
  We apply the Fermi-L\"owdin orbital (FLO) based SIC (FLOSIC) approach \cite{FERMI1,FERMI2,AAMOP} to examine the effect of SIC on polarizability
  as well as the IPs of these molecules. The presence of SIE leads
  to wrong asymptotic decay of the potential resulting in higher eigenvalues of the
  highest occupied molecular orbital (HOMO). This also affects 
  the asymptotic description of the electron density. SIC improves the potential
  in the asymptotic region which results in more tightly bound valence 
  electrons.  The absolute of the HOMO eigenvalues are significantly 
  improved in the SIC methods due to the improved description of valence 
  electrons in the asymptotic region. 
  The changes in the HOMO eigenvalues and polarizabilities due to SIC show how the SIE evolves with system size for these systems with $\pi$-electrons. 
  
  The orbital by orbital full SIC correction, however, can lead to 
  excessive corrections resulting in excessively corrected polarizability 
  values \cite{water_pol}. We show that excessive correction can be 
  removed using a locally-scaled SIC (LSIC) potential in the spirit of the LSIC 
  approach \cite{doi:10.1063/1.5129533}. Our earlier applications on water
  clusters showed that the LSIC polarizabilities are  in excellent agreement
  with reference CCSD values \cite{water_pol}.  The LSIC method 
  has also performed well for various properties of atoms and small molecules 
  such as ionization energies, total energies of atoms, atomization energies, 
  electron affinities, barrier heights of reactions, etc. \cite{doi:10.1063/1.5129533}. 
  The Perdew-Zunger orbital by orbital SIC has incorrect behavior of the 
  XC functionals at the uniform density limit where the semi-local (non-empirical)
  DFAs are designed to be accurate \cite{doi:10.1063/1.5090534}. Scaling down the SIC in 
  the many-electron density regions\cite{doi:10.1063/1.2176608,https://doi.org/10.1002/jcc.26168,doi:10.1063/5.0004738,doi:10.1063/1.5129533,sdSIC,D0CP06282K} improves the accuracy of the SIC. In this work,  
  we also examine the performance of this approach on the polyacenes which are conjugated systems. 
 
 In the following sections, we present the method, computational details, and the results and discussion.

\section{Method and computational details}

The SICs in the present work are obtained using the  
FLOs\cite{Leonard1982,Luken1982,Luken1984}. 
The FLO-SIC formalism \cite{FERMI1} provides an efficient approach to implement the orbital by orbital 
correction to DFA calculations. The PZSIC total energy is calculated as
\begin{equation}
    E^{DFA-SIC}=E^{DFA} - \sum_{i,\sigma}^{occ} \left\{ U[\rho_{i\sigma}]+E_{xc}[\rho_{i\sigma},0] \right \},
\end{equation}
where $U[\rho_{i\sigma}]$ and $E_{xc}[\rho_{i\sigma},0]$ are the self-Coulomb and self-XC energies of the $i^{th}$ orbital of spin index $\sigma$.
In the FLOSIC approach, a set of local FLOs\cite{Luken1982,Luken1984} is used to 
calculate the orbital-wise corrections. The Fermi orbitals are defined as
 \begin{equation}
     F_i(\vec {r})= \sum_j \frac {\psi_j(\vec{a_i}) \psi_j (\vec{r})}{\sqrt{\rho (\vec{a_i})}}.
 \end{equation}
Here, $\psi_j (\vec r) $ are the canonical Kohn-Sham orbitals of the system. The set of positions $\vec{a_i}$ 
are known as Fermi orbital descriptors (FODs) which determines the characteristics of the Fermi orbitals.
  The Fermi orbitals are further orthogonalized using the L\"owdin orthogonalization scheme \cite{Lowdin1956} 
  that yields the FLOs.   The use of Fermi orbitals leads to  unitarily invariant
  self-interaction corrected total energy. The minimization of the total SIC energy is carried out 
  through the optimization of the FODs \cite{FERMI2,AAMOP}.  In this work, 
  we refer to the PZSIC with FLO as FLOSIC.

We have chosen the LDA as parameterized in Ref. \cite{PW91} and the Perdew-Burke-Ernzerhof (PBE) GGA \cite{PBE} to 
XC functional for this study. These are the functionals that belong to 
the two lowest rungs of the Jacob ladder of functionals\cite{perdew2001jacob}.  Our earlier calculations on the 
polarizabilities of atoms and water clusters have shown that application of full PZSIC 
tends to overcorrect the polarizabilities bringing them lower than the reference 
values \cite{PhysRevA.100.012505,water_pol}. This overcorrection is reduced when a 
LSIC scheme is used. We use the following one-electron Hamiltonian 
with a scaling function $f(\vec{r})$ to scale the SIC potential \cite{water_pol}:

\begin{gather} \label{eq:ham}
     H_j = - \frac{1}{2} \nabla^2 + v_{ext}(\vec{r}) + \int \frac{\rho(\vec{r'})}{|\vec{r}-\vec{r'}|} d\vec{r'} + v_{XC}^{DFA}([\rho],\vec{r})  \\ \nonumber
     - f(\vec{r}) \left( 
     \int \frac{\rho_j(\vec{r'})}{|\vec{r}-\vec{r'}|} d\vec{r'}
     + v_{XC}^{DFA}([\rho_j],\vec{r}) \right) 
\end{gather}
where each term respectively denotes kinetic energy operator, external potential, 
Hartree potential, XC potential, and the last two terms in the bracket 
are the orbitalwise Hartree potential and XC potentials scaled by scaling factor
$f(\vec{r})$.
This Hamiltonian is similar in spirit to model XC potentials designed for accurate excitation energies or polarizability \cite{van1994exchange,van2001influence,gruning2002required,banerjee2007time,banerjee2008ab,schipper2000molecular}.
Any function that can identify the one-electron region could be used in Eq. (\ref{eq:ham}).
In this work, we use the scaling function defined as
$f(\vec{r})= z_{\sigma} (\vec {r})=\frac{\tau^W(\vec{r})}{\tau(\vec{r})}$
where $\tau^W(\vec{r}) = \frac {\lvert \nabla \rho(\vec{r})\rvert^2}{8 \rho({\vec{r})}} $ 
is the Weizs\"acker kinetic energy density and $\tau(\vec{r})$ is the kinetic energy
density. The function $z_\sigma$ is an iso-orbital indicator. At the uniform electron 
gas limit, the scaling factor becomes zero and approaches unity in the  one-electron density
region. Thus the scaling factor selects out the one-electron regions where the full SIC is 
applied and reduces the SIC in the many-electron region. In the uniform density limit 
where the semi-local DFAs are exact\cite{doi:10.1063/1.5090534}, SIC 
in the LSIC method vanishes. The LSIC method unlike PZSIC retains the exact behavior 
of the semi-local DFAs.

\begin{figure}
     \centering
     \includegraphics[width=0.8\textwidth]{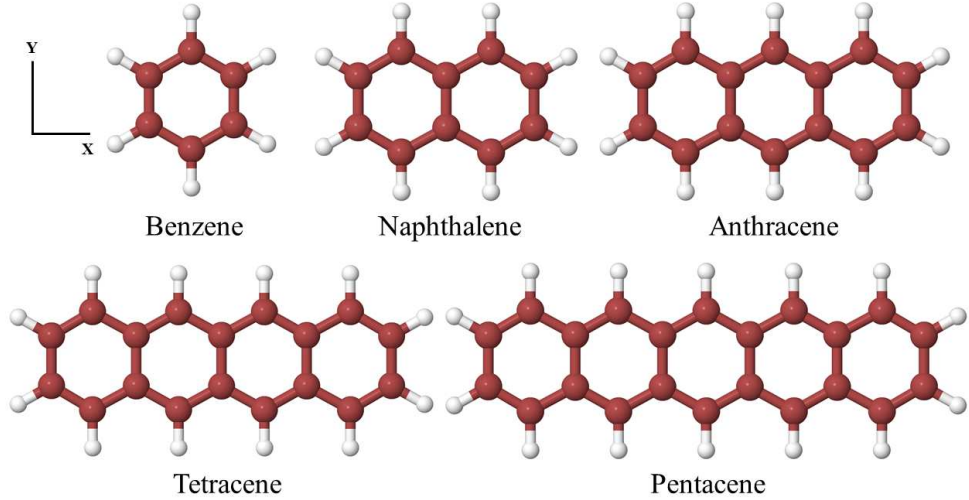} 
     \caption{Geometries of polyacenes.}
     \label{fig:geometry-poly}
\end{figure}

\begin{figure}
     \centering
     \includegraphics[width=0.8\textwidth]{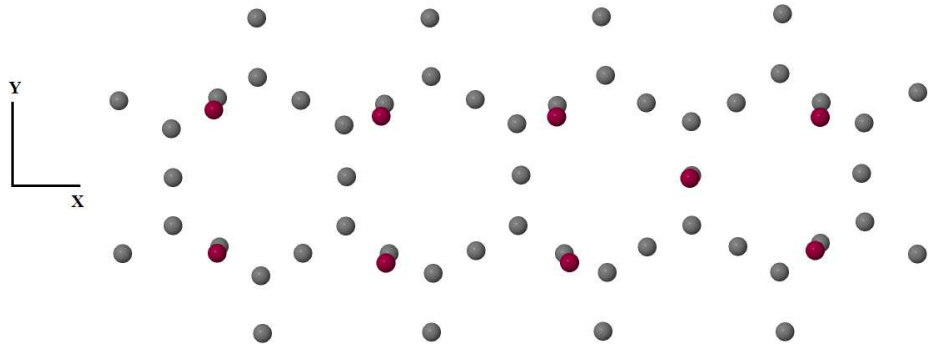} 
     \caption{Optimized FOD positions for tetracene (using LDA) in the absence of an electric field.}
     \label{fig:FOD-poly}
\end{figure}

All the calculations for both DFA and FLOSIC are done  using the recently developed FLOSIC code \cite{FERMI1,PhysRevA.95.052505,FLOSICcode}. We used the NRLMOL basis sets that contain 5, 4, and 3 contracted Gaussians of s, p, and d types respectively for C atom and 
4, 3, and 1 contracted functions of s, p, d types for H atom. We also used 
p and d type long-range single Gaussian polarization functions.
For electric field calculations, larger basis sets are used to interpret the response of polyacene to an external electric field correctly. Same basis sets are used for zero field calculations.
We apply an electric field strength of 0.005 a.u. along six directions ($\pm  x$, $\pm y$, $\pm z$) to compute the induced dipole moments and therefrom the polarizabilities.
Typically, the field amplitude that is proportional to geometrical progression\cite{C4CP00105B,doi:10.1021/acs.jctc.6b00360} may be used. In this 
work we used constant field amplitude as used in Ref.~\onlinecite{huzak2013benchmark}
for the CCSD(T) polarizabilities.
The field amplitude was chosen after testing various field 
amplitudes.
The FODs are optimized first for zero applied field with the two functionals  and these FOD’s are used later for calculations with the electric field.  
We point out that the local SIC potential can lead to symmetry breaking in the density which can manifest as a small but non-zero dipole moment in the polyacenes. 
To avoid further symmetry breaking the FODs were not optimized in the presence of the electric field. We also point out that the optimization of the FODs in the presence of electric field does not change the calculated polarizabilities \cite{small_mol} significantly while it adds to the computational costs. The methodology adopted here takes into account only the induced dipole moment.

\section{Results and discussion}
We used geometries of the polyacenes optimized with PBE functional.
The coordinates of the molecules can be found in Supplementary Materials.
The geometries and orientations of the polyacenes used in this work are shown in Fig. \ref{fig:geometry-poly}. All polyacenes are kept on the xy-plane with the molecular axis along the x-direction as shown in Fig. \ref{fig:geometry-poly}
which helps in determining the polarizability in the longitudinal direction. 
The optimized FOD positions for tetracene are shown in Fig. \ref{fig:FOD-poly}.  The resonance structure of the polyacenes leads to several floating FODs \cite{AAMOP,FLOSIC_porphyrin}. 

The application of a static electric field leads to induced dipole moments on the polyacenes which have zero dipole moments otherwise. The diagonal elements of the static dipole polarizability tensor calculated from the induced dipole moments for these molecules with LDA,  FLOSIC-LDA, PBE, FLOSIC-PBE, LSIC-LDA, and LSIC-PBE  are presented in Table \ref{table_pol1}. The calculated
polarizabilities are compared to available experimental and available calculated values with the 
CCSD method. Since the experimental values often contain the contributions from vibrational polarizability, we also include the  CCSD values for comparisons. 
 Hammond \textit{et al.} have carried out the calculation of polarizability of polyacenes up to hexacene using CCSD linear response method \cite{hammond2007dynamic} with the Sadlej pVTZ basis. Huzak \textit{et al.} have used the CCSD with perturbative triple excitations [CCSD(T)] with cc-pV$\infty$Z basis to calculate the polarizabilities of  naphthalene, anthracene, and tetracene with the finite field method \cite{huzak2013benchmark}. We include both these values in Table \ref{table_pol1} as well as experimental values. The CCSD and CCSD(T) average values are close for the smaller two molecules of the set and differ by 20-37 a.u. for anthracene and tetracene. They also differ in the opposite way from the experimental values for these two molecules.

\begin{longtable}{lccccc}
\caption{
The mean static dipole polarizability $\Bar{\alpha}$ and the polarizability components 
$\alpha_{xx}$, $\alpha_{yy}$, $\alpha_{zz}$  of polyacenes (in a.u.) within DFA, FLO-SIC, and LSIC methods.}
\label{table_pol1}\\
\hline
Molecules & Methods &$\alpha_{xx}$ &$\alpha_{yy}$&$\alpha_{zz}$&$\Bar{\alpha}$ \\
\hline
\endfirsthead

 \multicolumn{6}{c}%
 {\tablename\ \thetable\ -- \textit{Continued from previous page}} \\

\hline

Molecules & Methods &$\alpha_{xx}$ &$\alpha_{yy}$&$\alpha_{zz}$&$\Bar{\alpha}$ \\

\hline
\endhead
\hline \multicolumn{6}{r}{\textit{Continued on next page}}\\
\endfoot
\hline
\endlastfoot

 {\centering \textbf{Benzene}}
&	DFA-LDA	&	83.29	&	83.30	&	44.95	&	70.51	\\
&	DFA-PBE	&	82.98	&	82.99	&	44.59	&	70.19	\\
&	FLO-LDA	&	77.65	&	77.66	&	45.99	&	67.10	\\
 &	FLO-PBE	&	78.48	&	78.47	&	48.04	&	68.33	\\
 &	LSIC-LDA	&	79.95	&	79.95	&	42.33	&	67.41	\\
 &	LSIC-PBE	&	79.89	&	79.90	&	42.76	&	67.52	\\
&	\textbf{CCSD}$^x$	&	\textbf{80.57}	&	\textbf{80.57}	&	\textbf{44.66}	&	\textbf{68.60}	\\
&	\textbf{MP2$^z$}	&	\textbf{-}	&	\textbf{-}	&	\textbf{-}	&	\textbf{67.99}	\\
&	\textbf{Exp$^a$}	&	\textbf{-}	&	\textbf{-}	&	\textbf{-}	&	\textbf{67.79}	\\
 \colrule
  \colrule

  {\centering \textbf{Naphthalene}}
&	DFA-LDA	&	177.44	&	129.02	&	66.69	&	124.39	\\
&	DFA-PBE	&	176.95	&	128.51	&	66.16	&	123.88	\\
 &	FLO-LDA	&	157.04	&	115.59	&	64.93	&	112.52	\\
 &	FLO-PBE	&	158.33	&	116.67	&	67.22	&	114.07	\\
 &	LSIC-LDA	&	166.13	&	121.85	&	62.30	&	116.76	\\
 &	LSIC-PBE	&	166.71	&	121.66	&	62.70	& 116.69	\\

&	\textbf{CCSD}$^x$	&	\textbf{166.61}	&	\textbf{123.39}	&	\textbf{66.43}	&	\textbf{118.81}	\\
&	\textbf{CCSD(T)}$^y$	&	\textbf{162.20}	&	\textbf{120.00}	&	\textbf{65.30}	&	\textbf{115.83}	\\
&	\textbf{Exp$^b$}	&	\textbf{162}	&	\textbf{119.5}	&	\textbf{70.8}	&	\textbf{117.4}	\\
 \colrule

  {\centering \textbf{Anthracene}}
&	DFA-LDA	&	308.55	&	173.71	&	87.65	&	189.97	\\
&	DFA-PBE	&	307.55	&	172.99	&	87.00	&	189.18	\\
&	FLO-LDA	&	255.76	&	150.88	&	82.51	&	163.05	\\
&	FLO-PBE	&	264.74	&	155.50	&	88.10	&	169.45	\\
&	LSIC-LDA&	278.34	&	161.78	&	81.42	&	173.85	\\
&	LSIC-PBE	&	281.43	&	163.04	&	82.85	&	175.77	\\
&	\textbf{CCSD}$^x$	&	\textbf{281.6}	&	\textbf{166}	&	\textbf{87.58}	&	\textbf{178.39}	\\
&	\textbf{CCSD(T)}$^y$	&	\textbf{250.20}	&	\textbf{159.9}	&	\textbf{55.1}	&	\textbf{155.07}	\\
&	\textbf{Exp$^c$}	&	\textbf{302}	&	\textbf{153}	&	\textbf{70}	&	\textbf{175}	\\
 \colrule

 {\centering \textbf{Tetracene}}
 &	DFA-LDA	&	478.46	&	219.57	&	108.53	&	268.85	\\
&	DFA-PBE	&	476.64	&	218.82	&	107.73	&	267.73	\\
 &	FLO-LDA	&	396.01	&	194.77	& 106.11 &232.30		\\
 &	FLO-PBE	&	398.86	&	196.17	& 109.16	&	234.73	\\
 &	LSIC-LDA	&	431.10	&	206.31	&	102.25	&	246.56	\\
 &	LSIC-PBE	&	429.33	&	205.85	&	102.76	&	245.98	\\
&	\textbf{CCSD}$^x$	&	\textbf{423.83}	&	\textbf{209.77}	&	\textbf{108.61}	&	\textbf{247.40}	\\
&	\textbf{CCSD(T)}$^y$	&	\textbf{381.5}	&	\textbf{188.4}	&	\textbf{61.7}	&	\textbf{210.5}	\\
&	\textbf{Exp$^d$}	&	\textbf{-}	&	\textbf{-}	&	\textbf{-}	&	\textbf{217.8}	\\

 \colrule
  {\centering \textbf{Pentacene}}
&	DFA-LDA	&	686.81	&	266.58	&	129.32	&	360.90	\\
&	DFA-PBE	&	683.79	&	265.69	&	128.37	&	358.28	\\
&	FLO-LDA	&546.98		&233.29	&	124.98	&301.75		\\
&	FLO-PBE	&	552.74	&	235.93	&	128.88	&	305.85	\\
&	LSIC-LDA	&	606.01	&	248.80	&	121.63	&	325.49	\\
&	LSIC-PBE	&	604.26	&	248.66	&	122.34	&	325.09	\\
&	\textbf{CCSD}$^x$	&	\textbf{589.97}	&	\textbf{254.92}	&	\textbf{129.58}	&	\textbf{324.82}	\\
 
   \colrule

 \multicolumn{5}{l}{\small{$^x$ Sadlej pVTZ basis from reference[\onlinecite{hammond2007dynamic}]}}\\

 \multicolumn{5}{l}{\small{$^y$ Aug-cc-pV$\infty$Z basis from reference
 [\onlinecite{huzak2013benchmark}]}}\\
 \multicolumn{5}{l}{\small{$^z$ Reference
[\onlinecite{thakkar2015well}]}}\\
  \multicolumn{5}{l}{\small{$^a$ Reference [\onlinecite{HOHM2013282}]}}\\
 \multicolumn{5}{l}{\small{$^b$ Reference [\onlinecite{heitz1992measurement}]}}\\
 \multicolumn{5}{l}{\small{$^c$ Reference [\onlinecite{bendkowsky2007high}]}}\\
  \multicolumn{5}{l}{\small{$^d$ Reference
[\onlinecite{smith2004static}]}}\\

\end{longtable}

\begin{figure}[!ht]
 \centering

   \includegraphics[width=0.8\textwidth]{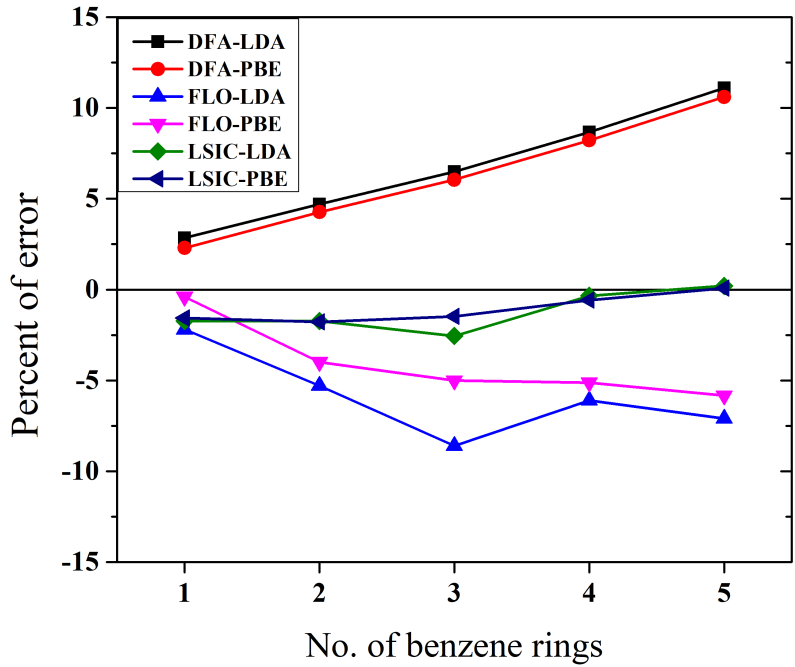}

  \caption{Percentage errors of calculated polarizability  of polyacene with respect to CCSD values reported in Ref. \cite {hammond2007dynamic}.}
  \label{fig:err}
\end{figure}

   In Fig. \ref{fig:err} we show the percentage error of our calculated polarizability values with respect to the CCSD values of Ref.~\cite{hammond2007dynamic}.
   This reference is chosen mostly for consistency for the whole set.
   As the system size increases the percentage deviation of polarizability with DFAs also increases. 
     The inclusion of one-electron SIC within the FLOSIC approach removes the overestimation but the correction is excessive. This trend was also seen in 
our previous results on polarizabilities of atoms, water clusters,  covalent and ionic molecules that showed that the LDA and PBE overestimate the polarizabilities compared to the reference CCSD/CCSD(T) values \cite{PhysRevA.100.012505,water_pol,small_mol}.
The errors in FLOSIC polarizability values also tend to increase with size but the increase is not systematic. The polarizability of the aromatic polyacenes does not linearly   increase with the number of rings as evident from the experimental and CCSD/CCSD(T) values.
 Yang and co-workers \cite{Weitao_pol} 
 have tested the polarizability of the H$_2$ molecular chain with one-electron SIE-free 
functionals B05 \cite{doi:10.1063/1.1844493} and MCY \cite{doi:10.1063/1.2179072}. 
The results show that the removal of one-electron SIE reduces the polarizabilities but the values still had a trend similar to that shown by LDA in that the polarizabilities were overestimated compared to reference MP2 results and the overestimation increased with system size.
The FLOSIC approach corrects for the one-electron SIE but
 the FLOSIC polarizabilities are consistently lower than the reference values, and the errors of FLOSIC values do not show a  trend similar to DFAs. The maximum error for  FLOSIC-LDA occurs for anthracene and decreases for the larger molecules. On the other hand, for FLOSIC-PBE absolute percentage errors increase with size but the increase rate is low for the larger systems.

In the LSIC approach, the SIC potential is locally scaled such that in the one-electron regions
full SIC is applied but it is scaled down in the many-electron regions. 
In the polyacenes, the one-electron regions are around the hydrogens and the core regions 
of the carbons atoms. The iso-orbital indicator becomes small in the interatomic regions 
on the C-C bonds and also on the C-H bonds and as a result, the SIC potential is scaled down in these regions.
The LSIC polarizability values lie between FLOSIC and DFA values and are in excellent 
agreement with the reference CCSD values (Ref. \cite{hammond2007dynamic}) with the mean 
absolute percentage errors of 1.3\% and 1.1\% for LSIC-LDA and LSIC-PBE, respectively. 
The LSIC values with LDA and PBE are close except for the anthracene. The LSIC percentage 
errors start with a negative value for benzene and increase with system size. The polarizability 
components present a clearer picture of how the density functional polarizabilities  compare 
with CCSD values. For $\alpha_{zz}$ the DFA and FLOSIC-DFA polarizability values are in good
agreement with CCSD values. The DFA overestimation and FLOSIC underestimation worsen for 
the other two dimensions where the system sizes are larger.  
Overall, the LSIC-PBE performs best for the calculation of polarizability of these systems in the size range considered here. Similar conclusions were also reached for the water cluster polarizability too \cite{water_pol}.

\begin{figure}
    \centering
    \includegraphics[width=0.9\columnwidth]{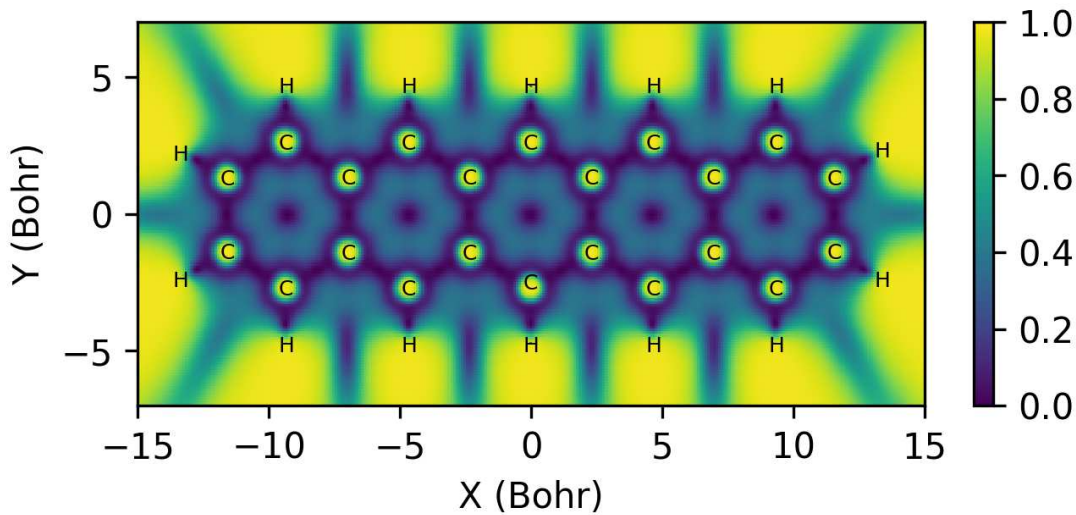}
    \caption{A cross sectional plot of the iso-orbital indicator $z_\sigma(\vec{r})$ for pentacene obtained at the PBE level of theory. The iso-orbital indicator finds the regions around C atoms to be many-electrons like. }
    \label{fig:iso-orb}
\end{figure}

\begin{table}[!tbp] 
      \caption{The negative of HOMO eigenvalues of the polyacenes (in eV) using the DFA, FLO-SIC, and LSIC approaches compared with CCSD(T)-cc-PVDG and experimental results.}
      \label{table_IP}

\begin{tabular*}{0.50\textwidth}{@{\extracolsep{\fill}}ccccccccc}
\colrule
No. of &\multicolumn{2}{c}{DFA} & \multicolumn{2}{c}{FLO-SIC}& \multicolumn{2}{c}{LSIC}& CCSD$^x$ & Exp$^y$
\\\cline{2-3} \cline{4-5}  \cline{6-7} 
  ring & LDA & PBE & LDA & PBE & LDA & PBE & & \\  \hline
1	&	6.55	&	6.35	&	9.32	&	8.80	&	9.48	&	9.11	&	9.17	&	9.2437	\\
2	&	5.69	&	5.48	&	8.89	&	8.35	&	8.83	&	8.46	&	7.96	&	8.144	\\
3	&	5.19	&	4.98	&	8.73	&	7.86	&	8.44	&	7.88	&	7.19	&	7.439	\\
4	&	4.88	&	4.66	&	7.84	&	7.32	&	7.77	&	7.42	&	6.66	&	6.97	\\
5	&	4.66	&	4.44	&	7.63	&	7.06	& 7.53		&	7.15	&	6.28	&	6.63	\\
 MAE$^x$	&	2.06	&	2.27	&	1.03	&	0.57	&	0.96	&	0.58	&		&		\\
 MAE$^y$	&	2.29	&	2.50	&	0.80	&	0.37	&	0.72	& 0.37		&		&		\\

\colrule
\multicolumn{8}{l}{\small{$^x$ Reference [\onlinecite{deleuze2003benchmark}]}}\\ 
\multicolumn{8}{l}{\small{$^y$ Reference [\onlinecite{linstrom2001nist}]}}\\
\end{tabular*}      
\end{table}

We also compared eigenvalues of the HOMO  
that mimics the negative of the IP with CCSD(T)  and 
experimental IPs in Table \ref{table_IP}. 
For the exact energy functional, the absolute of HOMO eigenvalue equals
the vertical IP\cite{perdew1982density,levy1984exact,perdew1997comment,PhysRevB.60.4545}.
The CCSD(T) values are calculated from the $\Delta$-SCF method with the cc-pVDZ basis \cite{deleuze2003benchmark}.
  The  SIE in DFAs leads to faster asymptotic decay of the potential resulting in higher 
  eigenvalues for the HOMO and consequently lower IPs.  
The mean absolute errors in DFA, FLOSIC-DFA, and LSIC-DFA show that the application 
of SIC reduces the errors in DFA significantly.  We also note that the LSIC-PBE HOMO eigenvalues are 
slightly lower than those with FLOSIC-PBE. Only for benzene, LSIC-LDA HOMO eigenvalue is lower than 
that of FLOSIC-LDA.  However, the mean absolute error in HOMO eigenvalues is lower with LSIC than 
with FLOSIC for LDA whereas it remains nearly the same for PBE. 
Since LSIC scales the SIC potentials in the many-electron regions while keeping the same potential in one-electron regions,  SIC potential in many-electron regions
in these molecules is likely positive.
A plot of the iso-orbital indicator $z_{\sigma}(\vec{r})$ on the plan of the molecule is shown in Fig.~\ref{fig:iso-orb}. This plot
shows the regions around the C atoms as the many-electron like where the SIC potential is scaled.
The plot also demonstrates that the iso-orbital indicator can be small in regions far from the molecules that are not true many-electron like regions as discussed by Schmidth \textit{et al.} \cite{schmidt2014one}.  
Overall, the SIC corrections to HOMO eigenvalues are negative for all the molecules studied here. 
We also point out that while the LSIC-LDA and LSIC-Coulomb energy functionals have 
the same gauge, this is not the case for the LSIC-GGA and LSIC-meta-GGA functionals\cite{sdSIC}.
However, our earlier calculations \cite{water_pol,sdSIC,small_mol} and the results 
reported here show that the LSIC-PBE performs exceedingly well despite the formal theoretical 
shortcomings. Overall, the absolute HOMO eigenvalues with FLOSIC-PBE and LSIC-PBE are 
in very good agreement with both theoretical and experimental reference values.

In summary, we have applied the one-electron SIC through the FLOSIC and LSIC methods to calculate the static 
dipole polarizability of the lowest five polyacenes -- from benzene to pentacene. The FLOSIC performs an 
orbital by orbital SIC. The overestimating trend of LDA and PBE is corrected by 
FLOSIC (PZSIC with FLOs) but the correction is excessive. This trend of FLOSIC has been noted for different
types of systems -- from atoms and anions, water clusters, and covalent and ionic 
molecules\cite{PhysRevA.100.012505,small_mol}
The errors in polarizability with FLOSIC does not systematically increase with the size of the polyacenes for LDA.
We also find that local scaling of SIC potential reduces the overcorrection and leads to  polarizability
values that are in excellent agreement with CCSD/CCSD(T) values and also with the experiment. The FLOSIC and 
LSIC corrections to the HOMO eigenvalues also show good agreement with CCSD values and experiment. The 
need for SIC in  calculations of polarizability for long chain-like has been demonstrated earlier for 
model systems and polymers. This work shows the DFA calculations with LSIC can produce results that are
in excellent agreement with CCSD for these systems.

\section*{Supplementary material}
See supplementary material for the coordinates of the molecules studied in this manuscript.

\section*{Dedication}
This paper honors Prof. Anjali Kshirsagar, who is an outstanding teacher of computational 
condensed matter and chemical physics. Prof. Kshirsagar has mentored and inspired many students,
including the two senior authors, to pursue careers in chemical physics during her career
as a Professor at Savitribai Phule Pune University. 

\section*{Acknowledgment}
This work was supported by the U.S. Department of Energy, Office of Science, Office of Basic Energy Sciences, as part of the Computational Chemical Sciences Program under Award No. DE-SC0018331. S.A. was partly supported by the U.S. Department of Energy, Office of Science, Office of Basic Energy Sciences, under Award No. DE-SC0002168. Support for computational time at the Texas Advanced Computing Center through NSF (Grant No. TG-DMR090071) and at NERSC is gratefully acknowledged.

\section*{Data Availability}
The data that support the findings of this study are available within the article.

\bibliography{combined_refs}
\end{document}


{\let\vfil\relax{\let\clearpage\relax\maketitle}}

\begin{table}
\caption{\label{tab:stab1}Cartesian coordinates for polyacenes from benzene to pentacene}
\begin{tabular*}{\columnwidth}{@{\extracolsep{\fill}}crrrr}
\toprule
\vspace{5mm}	
Benzene	&	&	 & 	&\\

12	&		&		\\
C   &  0.0000	&	1.3858	&	0.0000	\\
C   &  0.0000	&	-1.3858 &	0.0000	\\
C   &  1.2006	&	0.6934	&	0.0000	\\
C  &  -1.2006	&	0.6934	&	0.0000	\\
C  &   1.2006	&	-0.6934 &	0.0000	\\
C   & -1.2006	&	-0.6934 &	0.0000	\\
H    & 0.0000	&	2.4797	&	0.0000	\\
H    & 0.0000	&	-2.4797 &	0.0000	\\
H   &  2.1472	&	1.2405	&	0.0000	\\
H   & -2.1472	&	1.2405	&	0.0000	\\
H   &  2.1472	&	-1.2405 &	0.0000	\\
H    &-2.1472	&	-1.2405 &	0.0000	\\
\end{tabular*}
\end{table}
\begin{table}
\begin{tabular*}{\columnwidth}{@{\extracolsep{\fill}}crrrr}
\toprule

\vspace{5mm}

Naphthalene	&		&		&		\\

18	&		&		&		\\
C	&	1.2195	&	0.7086	&	0.0000	\\
C	&	1.2195	&  -0.7086	&	0.0000	\\
C	&	0.0282	&	1.4049	&	0.0000	\\
C	&	0.0282	&  -1.4049	&	0.0000	\\
C	&	-1.2166	&	0.7182	&	0.0000	\\
C	&	-1.2166	&	-0.7182	&	0.0000	\\
C	&	-3.6530	&	0.7089	&	0.0000	\\
C	&	-3.6530	&	-0.7089	&	0.0000	\\
C	&	-2.4609	&	1.4053	&	0.0000	\\
C	&	-2.4609	&	-1.4053	&	0.0000	\\
H	&	0.0226	&	2.4972	&	0.0000	\\
H	&	0.0226	&	-2.4972	&	0.0000	\\
H	&	-2.4542	&	2.4977	&	0.0000	\\
H	&	-2.4542	&	-2.4977	&	0.0000	\\
H	&	2.1678	&	1.2486	&	0.0000	\\
H	&	2.1678	&	-1.2486	&	0.0000	\\
H	&	-4.6012	&	1.2494	&	0.0000	\\
H	&	-4.6012	&	-1.2494	&	0.0000	\\
\end{tabular*}
\end{table}
\begin{table}
\begin{tabular*}{\columnwidth}{@{\extracolsep{\fill}}crrrr}
\toprule

\vspace{5mm}

Anthracene	&		&		&		\\
24	&		&		&		\\
C	&	1.2250	&	0.7236	&	0.0000	\\
C	&	1.2250	&  -0.7236	&	0.0000	\\
C	&	-0.0001	&	1.4058	&	0.0000	\\
C	&	-0.0001	&  -1.4058	&  -0.0000	\\
C	&	3.6639	&	0.7125	&	0.0000	\\
C	&	3.6639	&	-0.7125	&	0.0000	\\
C	&	2.4786	&	1.4079	&	0.0000	\\
C	&	2.4786	&	-1.4079	&	0.0000	\\
C	&	-1.2252	&	0.7235	&	0.0000	\\
C	&	-1.2252	&	-0.7235	&	0.0000	\\
C	&	-3.6642	&	0.7124	&	0.0000	\\
C	&	-3.6642	&	-0.7124	&	0.0000	\\
C	&	-2.4788	&	1.4077	&	0.0000	\\
C	&	-2.4788	&	-1.4077	&	0.0000	\\
H	&	-0.0001	&	2.4982	&	0.0000	\\
H	&	-0.0001	&	-2.4982	&  -0.0000	\\
H	&	2.4756	&	2.4994	&	-0.0000	\\
H	&	2.4756	&	-2.4994	&	-0.0000	\\
H	&	-2.4758	&	2.4993	&	0.0000	\\
H	&	-2.4758	&	-2.4993	&	0.0000	\\
H	&	-4.6136	&	1.2494	&	0.0000	\\
H	&	-4.6136	&	-1.2494	&	0.0000	\\
H	&	4.6133	&	1.2494	&	0.0000	\\
H	&	4.6133	&	-1.2494	&	0.0000	\\

\end{tabular*}
 \end{table}
 
 \begin{table}
\begin{tabular*}{\columnwidth}{@{\extracolsep{\fill}}crrrr}
\toprule

\vspace{5mm}

Tetracene	&		&		&		\\
30	&		&		&		\\
C	&	1.2379	&	0.7266	&	-0.0000	\\
C	&	1.2379	&	-0.7266	&	-0.0000	\\
C	&	0.0198	&	1.4085	&	-0.0000	\\
C	&	0.0198	&	-1.4085	&	-0.0000	\\
C	&	3.6775	&	0.7141	&	-0.0000	\\
C	&	3.6775	&	-0.7141	&	-0.0000	\\
C	&	2.4953	&	1.4097	&	-0.0000	\\
C	&	2.4953	&	-1.4097	&	-0.0000	\\
C	&	-1.2156	&	0.7275	&	-0.0000	\\
C	&	-1.2156	&	-0.7275	&	-0.0000	\\
C	&	-3.6687	&	0.7266	&	-0.0000	\\
C	&	-3.6687	&	-0.7266	&	-0.0000	\\
C	&	-2.4505	&	1.4087	&	-0.0000	\\
C	&	-2.4505	&	-1.4087	&	-0.0000	\\
C	&	-4.9265	&	1.4092	&	0.0000	\\
C	&	-4.9265	&	-1.4092	&	0.0000	\\
C	&	-6.1091	&	0.7141	&	0.0000	\\
C	&	-6.1091	&	-0.7141	&	0.0000	\\
H	&	0.0203	&	2.5012	&	-0.0000	\\
H	&	0.0203	&	-2.5012	&	-0.0000	\\
H	&	2.4920	&	2.5013	&	0.0000	\\
H	&	2.4920	&	-2.5013	&	-0.0000	\\
H	&	-2.4468	&	2.5013	&	-0.0000	\\
H	&	-2.4468	&	-2.5013	&	-0.0000	\\
H	&	-4.9212	&	2.5005	&	-0.0000	\\
H	&	-4.9212	&	-2.5005	&	-0.0000	\\
H	&	-7.0594	&	1.2496	&	0.0000	\\
H	&	-7.0594	&	-1.2496	&	0.0000	\\
H	&	4.6269	&	1.2501	&	-0.0000	\\
H	&	4.6269	&	-1.2501	&	-0.0000	\\
\bottomrule
\end{tabular*}
 \end{table}

 \begin{table}
\begin{tabular*}{\columnwidth}{@{\extracolsep{\fill}}crrrr}
\toprule

\vspace{5mm}

Pentacene	&		&		&		\\
36	&		&		&		\\
C	&	1.2274	&	0.7293	&	-0.0000	\\
C	&	1.2274	&	-0.7293	&	-0.0000	\\
C	&	0.0001	&	1.4097	&	-0.0000	\\
C	&	0.0001	&	-1.4097	&	-0.0000	\\
C	&	3.6816	&	0.7278	&	-0.0000	\\
C	&	3.6816	&	-0.7279	&	0.0000	\\
C	&	2.4667	&	1.4094	&	-0.0000	\\
C	&	2.4667	&	-1.4094	&	-0.0000	\\
C	&	4.9416	&	1.4095	&	-0.0000	\\
C	&	4.9417	&	-1.4095	&	0.0000	\\
C	&	6.1239	&	0.7151	&	-0.0000	\\
C	&	6.1238	&	-0.7151	&	0.0000	\\
C	&	-1.2276	&	0.7293	&	-0.0000	\\
C	&	-1.2276	&	-0.7293	&	-0.0000	\\
C	&	-3.6818	&	0.7280	&	0.0000	\\
C	&	-3.6817	&	-0.7280	&	0.0000	\\
C	&	-2.4668	&	1.4094	&	0.0000	\\
C	&	-2.4669	&	-1.4094	&	0.0000	\\
C	&	-4.9419	&	1.4093	&	0.0000	\\
C	&	-4.9421	&	-1.4093	&	0.0000	\\
C	&	-6.1240	&	0.7152	&	0.0000	\\
C	&	-6.1239	&	-0.7152	&	0.0000	\\
H	&	0.0003	&	2.5017	&	-0.0000	\\
H	&	0.0003	&	-2.5017	&	-0.0000	\\
H	&	2.4597	&	2.5014	&	0.0000	\\
H	&	2.4597	&	-2.5014	&	0.0000	\\
H	&	4.9351	&	2.5004	&	-0.0000	\\
H	&	4.9351	&	-2.5004	&	0.0000	\\
H	&	7.0746	&	1.2498	&	-0.0000	\\
H	&	7.0746	&	-1.2498	&	-0.0000	\\
H	&	-2.4601	&	2.5013	&	-0.0000	\\
H	&	-2.4601	&	-2.5013	&	-0.0000	\\
H	&	-4.9354	&	2.5004	&	0.0000	\\
H	&	-4.9354	&	-2.5004	&	-0.0000	\\
H	&	-7.0749	&	1.2498	&	0.0000	\\
H	&	-7.0749	&	-1.2498	&	0.0000	\\
\bottomrule
\end{tabular*}
 \end{table}